\documentclass{PoS}

\title{Towards small-$x$ resummed parton distribution functions}
\ShortTitle{Towards small-$x$ resummed parton distribution functions}

\author{\speaker{Luca Rottoli}\\
        Rudolf Peierls Centre for Theoretical Physics, 1 Keble Road, University of Oxford, OX1 3NP Oxford, United Kingdom\\
        E-mail: \email{luca.rottoli@physics.ox.ac.uk}}

\author{Marco Bonvini\\
        Dipartimento di Fisica, Sapienza Universit\`a di Roma and INFN, Sezione di Roma, Piazzale Aldo Moro 5, 00185 Roma, Italy\\
        E-mail: \email{marco.bonvini@roma1.infn.it}}

\abstract
{We present preliminary results for fits of parton distribution functions (PDFs)
  which include the resummation of small-$x$ logarithms at NLLx accuracy, performed in the NNPDF framework.
  We observe an improvement in the description of DIS data at small values of Bjorken $x$ when resummation effects are included.
  The improvement is more marked when comparing NNLO+NLLx fits to NNLO ones,
  and is particularly noticeable for small-$x$ and small-$Q^2$  HERA inclusive structure functions.
  The main effect of the resummation is an enhancement of the gluon and singlet PDFs at small-$x$, which persists at high scales.
}

\FullConference{XXV International Workshop on Deep-Inelastic Scattering and Related Subjects\\
		3-7 April 2017\\
		University of Birmingham, UK}

\usepackage{url}
\usepackage{cite}
\usepackage{blindtext}
\usepackage{booktabs}

\begin{document}

Global parton distribution function (PDF) sets~\cite{Jimenez-Delgado:2014twa,Harland-Lang:2014zoa,Dulat:2015mca,Accardi:2016qay,Alekhin:2017kpj,Ball:2017nwa} are extracted from a variety of data collected over the years in different experiments.
%
%
PDFs depend on a dimensionful scale $Q$, the hard scale of the process, and a dimensionless scale $x$, which represents the proton momentum fraction carried by the parton.
The extraction of precise PDFs therefore depends not only on the precision of the available data, but also, crucially, on an accurate theoretical description of the physical observables in the ($x$,$Q^2$) range probed by the experiments.

Currently, PDFs are determined using fixed-order theory:
partonic cross sections are included up to next-to-next-to-leading order (NNLO) accuracy and PDF evolution is computed using splitting functions up to NNLO. 
A fixed-order perturbative description, however, may not be accurate enough to reliably describe all the processes included in PDF fits. 
For instance, at large $x$ threshold effects need to be taken into account~\cite{Corcella:2005us,Bonvini:2015ira}.
Furthermore,
it has been observed~\cite{Caola:2009iy,Caola:2010cy,Abramowicz:2015mha} that some tensions appear
in the description of low-$x$, low $Q^2$ data from the HERA collaboration~\cite{Abramowicz:2015mha}. 

In the latter kinematic region one should indeed supplement the fixed-order description with the resummation of a class of logarithms of $x$, which become large in the small-$x$, or high-energy, regime. 
In $\overline{\rm MS}$-like schemes, these logarithms appear both in the splitting functions and in the partonic cross sections. 
The formalism for resumming these high-energy logarithms has been developed in the past thirty years by several groups~\cite{Catani:1990xk,Catani:1990eg,Collins:1991ty,Catani:1993ww,Catani:1993rn,Catani:1994sq,Salam:1998tj,Ciafaloni:1999yw,Ciafaloni:2003rd,Ciafaloni:2007gf,Ball:2001pq,Ball:1995vc,Ball:1997vf,Altarelli:2001ji,Altarelli:2003hk,Altarelli:2005ni,Altarelli:2008aj,Rojo:2009us,Ball:2007ra,Caola:2010kv,Thorne:1999sg,Thorne:1999rb,Thorne:2001nr,White:2006yh}.
Despite the wealth of theoretical computations and developments, the number of phenomenological studies has been rather limited.
For these, crucially, PDF sets extracted making consistent use of small-$x$ resummation are needed, as high-energy logarithms affect mostly PDF evolution.
So far, only one PDF fit~\cite{White:2006yh} with the inclusion of high-energy resummation has been produced.

The main reason for the lack of PDF fits and phenomenological analyses is related to the rather involved technical details of this particular resummation.
%
Recently, the Altarelli-Ball-Forte approach to high-energy resummation~\cite{Ball:1995vc,Ball:1997vf,Altarelli:2001ji,Altarelli:2003hk,Altarelli:2005ni,Altarelli:2008aj,Rojo:2009us} has been revived and improved in~\cite{Bonvini:2016wki} to facilitate a systematic inclusion of high-energy resummation in different processes.
Most importantly, the non trivial results of high-energy resummation have been made publicly available through the code \texttt{HELL}~\cite{HELL}, which delivers
resummed splitting functions and partonic coefficient functions through a fast interface.
The \texttt{HELL} code has been interfaced to the evolution library \texttt{APFEL}~\cite{Bertone:2013vaa}, thus providing a framework for a systematic inclusion of small-$x$ resummation in PDF fits. 

Recently, the work of~\cite{Bonvini:2016wki} has been further developed~\cite{Bonvini:2017} to achieve two important results:
\begin{enumerate}
\item Resummed splitting functions, known up to next-to-leading-logarithmic (NLLx) accuracy,
  can now be matched to NNLO (previously only LO+LLx and NLO+NLLx results were available).
  This allows to fit for the first time a NNLO PDF set which includes small-$x$ resummation.
\item A complete description of deep-inelastic scattering (DIS) data can now be achieved,
  thanks to the inclusion of heavy-quark mass corrections to the resummed DIS coefficient functions.
  The new version of \texttt{HELL} thus implements the resummation of collinear mass logarithms 
  in the context of the FONLL~\cite{Forte:2010ta,Ball:2015tna,Ball:2015dpa} variable-flavour number scheme (VFNS) at the small-$x$ resummed level.
  As a byproduct, the VFNS matching conditions, which relate the PDF in two schemes with different number of active flavours,
  are resummed as well. This generalizes a previous work~\cite{White:2006xv} to $\overline{\rm MS}$-like schemes.
\end{enumerate}

\begin{figure}[t]
\centering
  \includegraphics[width=0.495\textwidth, page=1]{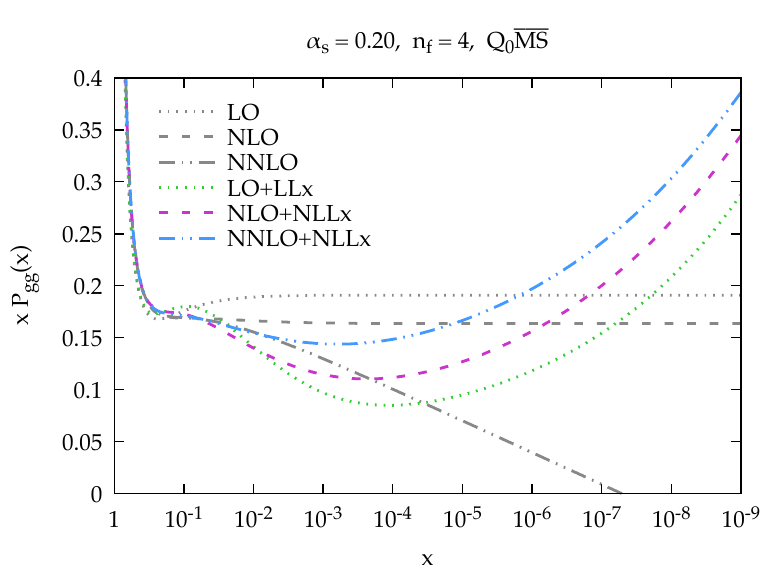}
  \includegraphics[width=0.495\textwidth, page=3]{plots/plot_P_nf4_paper_noband.pdf}
  \caption{Comparison of the fixed-order and matched gluon-gluon
    $xP_{gg}(x,\alpha_s)$ (left) and the quark-gluon $xP_{qg}(x,\alpha_s)$ (right) splitting functions~\cite{Bonvini:2017}.}
  \label{fig:splittingfunctions}
\end{figure}

The importance of having matched NNLO+NLLx splitting functions can be appreciated by looking at fig.~\ref{fig:splittingfunctions}. 
Whereas at LO and NLO $xP_{gg}$ (and $xP_{gq}$) splitting function does not grow logarithmically at small $x$, at NNLO it decreases as $\ln \frac{1}{x}$ with a negative coefficient, thus diverting from the resummed prediction, which instead rises in the small-$x$ region (thus producing the well-known `dip' structure~\cite{Ciafaloni:2003kd}).
The NNLO result, then, starts differing significantly from the NNLO+NLLx result for $x\lesssim 10^{-3}$. On the other hand, the NLO
is very close to (our best prediction of) the all-order result, NNLO+NLLx, for all $x\gtrsim10^{-6}$.
Thus, we expect NLO theory to describe gluon evolution accidentally well for a large range of $x$, while NNLO theory is expected to
degrade the theoretical description for $x\lesssim10^{-3}$.
NLO+NLLx is only marginally better than NLO, also because it differs quite significantly from the more precise NNLO+NLLx in intermediate regions of $x$.
Therefore, only NNLO+NLLx represents a substantial improvement in the theoretical description of the small-$x$ evolution, the difference being more substantial when compared to fixed NNLO theory.

The possibility of having a consistent matching to NNLO, as well as the access to the resummed massive DIS coefficient functions, allow to perform for the first time a NNLO+NLLx PDF fit to DIS data.
In this contribution, we present \emph{preliminary} results~\cite{NNPDFsx} for a fit of PDFs in the NNPDF framework, using a DIS-only dataset.
The fits are largely based on the same settings used in the recently released NNPDF3.1 set~\cite{Ball:2017nwa};
the total charm PDF is fitted along the light-quark and gluon PDFs~\cite{Ball:2016neh}. 
We produce fits at NLO, NLO+NLLx, NNLO, and NNLO+NLLx accuracy.

\begin{table}[t]
\begin{center}
{\small
\begin{tabular}{ c  c  c  c c }
  Experiment & NLO $\chi^2$ & NLO+NLLx $\chi^2$  & NNLO $\chi^2$  & NNLO+NLLx $\chi^2$   \\
\midrule
 NMC                   & 1.29  & 1.28 & 1.27 & 1.28 \\
 SLAC                  & 0.96  & 1.00 & 0.86 & 0.86 \\
 BCDMS                 & 1.19  & 1.20 & 1.20 & 1.20 \\
 CHORUS                & 1.01  & 1.00 & 1.00 & 0.99 \\
 NuTeV dimuon          & 0.59  & 0.56 & 0.55 & 0.55 \\
 HERA I+II             & 1.14  & 1.13 & 1.19 & 1.13 \\
 HERA $\sigma_c^{NC}$  & 1.59  & 2.15 & 1.26 & 1.37 \\
 HERA $F_2^b$          & 1.05  & 1.02 & 1.13 & 1.07 \\
 \midrule
 total                 & 1.117 & 1.118 & 1.126 & 1.104 \\
\end{tabular}}
\caption{The values of $\chi^2/N_{\textrm{dat}}$ for the NLO, NLO+NLLx, NNLO, and NNLO+NLLx fit to DIS data. 
}
\label{tab:chi2}
\end{center}
\end{table}

The results of the fits are collected in table~\ref{tab:chi2}. 
By inspecting the values of the total $\chi^2/N_{\textrm{dat}}$ we observe that the NNLO+NLLx fit has the lowest $\chi^2$ and significantly improves with respect to the NNLO fit, which on the contrary has the highest $\chi^2$.  
On the other hand, the NLO and the NLO+NLLx fits have essentially the same $\chi^2$, the difference being within statistical fluctuations.
The improvement of the total $\chi^2$ at NNLO+NLLx is mostly due to a better description of the HERA I+II dataset, which includes more than one third of the DIS datapoints included in the fit. 
Indeed, the $\chi^2$ to HERA data is $1.19$ at NNLO, and decreases to $1.13$ when high-energy resummation is included. 

\begin{figure}[t]
\centering
  \includegraphics[width=0.495\textwidth, page=1]{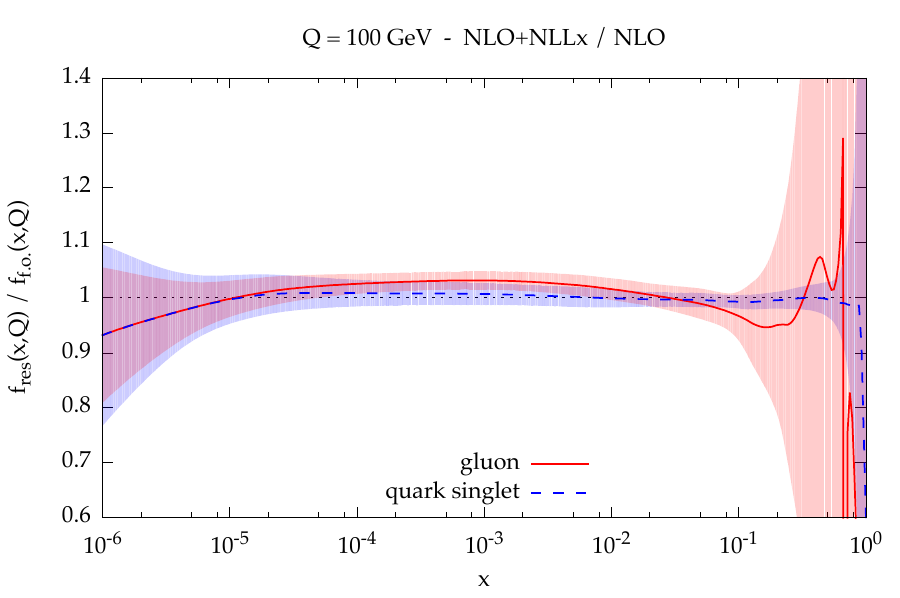}
  \includegraphics[width=0.495\textwidth, page=1]{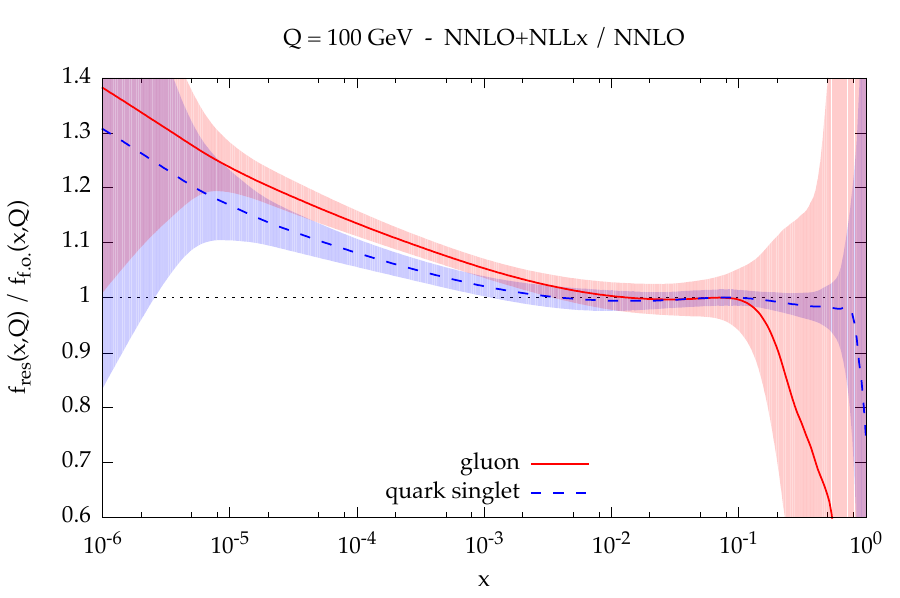}
  \caption{\small Ratio of the resummed gluon and quark singlet PDFs to the fixed order PDFs at 100 GeV. Left plot: NLO+NLLx/NLO; right plot: NNLO+NLLx/NNLO.   %
  }
  \label{fig:pdfs}
\end{figure}

The effect of resummation on the PDFs is shown in fig.~\ref{fig:pdfs}, where we show the ratio of the resummed gluon and quark singlet PDF to their fixed order counterpart at $100$ GeV, at NLO and at NNLO. 
At NLO the resummed quark singlet PDF is very similar to the fixed order PDF: differences are within the one sigma level.
%
The effect is much larger at NNLO: both the gluon and quark singlet PDFs are enhanced for $x\lesssim 0.01$ if resummation is included, up to $20\%$ for values of $x\sim10^{-5}$,
with significance of four-five sigmas.

The inclusion of high-energy resummation has relevant consequences on phenomenology at hadron colliders. 
For instance, we can assess the effect of resummed PDFs on the total cross section for Higgs production in gluon fusion at the LHC. 
The inclusive N$^3$LO cross section (computed with the public code \texttt{ggHiggs}~\cite{ggHiggs,Ball:2013bra,Bonvini:2016frm})
is 47.2~pb with the preliminary NNLO PDFs and 48.1~pb with the preliminary NNLO+NLLx PDFs;
the $\mathcal{O}(1\%)$ difference is comparable to the $1.2\%$ 
uncertainty associated to the lack of knowledge of N$^3$LO PDFs~\cite{deFlorian:2016spz}.
To have a consistent picture, however, small-$x$ resummation should be included in the coefficient functions for Higgs production~\cite{Bonvini:2017higgs}. 
%

Whilst all the ingredients for a NNLO+NLLx fit to DIS data are now available,
fitting PDFs from a DIS dataset would not make them competitive with global fits, which feature smaller PDF uncertainties over a wider range of $x$.
General-purpose small-$x$ resummed PDFs necessarily need to be determined from a variety of processes on top of DIS, such as Drell-Yan or jet production.
To this end, one should use resummed coefficient functions for all the other processes used for the extraction of PDFs, which requires some further theoretical and code development.
Nevertheless, a global fit with small-$x$ resummation is possible if a conservative cut is applied to non-DIS data in order to exclude from the fit datapoints which probe PDFs at small-$x$  (and would therefore be sensitive to small-$x$ logarithms), pending the inclusion of resummation effects in the coefficient functions for these processes. 
In this way, the resulting PDF set will faithfully describe the small-$x$ region, and have an accuracy competitive to fixed-order mainstream global fits.
A first global NNLO+NLLx fit will be the subject of a forthcoming paper~\cite{NNPDFsx}, of which the results presented in these proceedings represent a preliminary version.

\bibliography{br_DIS.bib}

\end{document}